\documentclass[aps,prd,showpacs,superscriptaddress,letterpaper,preprintnumbers,asmmath,amssymb]{revtex4}
\usepackage{amssymb,amsmath,amsbsy}
\usepackage{mathrsfs}
\usepackage[dvips]{graphics}
\usepackage{epsfig}
\usepackage{feynmf}

\def\be{\begin{eqnarray}}
\def\ee{\end{eqnarray}}
\def\bea{\begin{eqnarray*}}
\def\eea{\end{eqnarray*}}

\def\centeron#1#2{{\setbox0=\hbox{#1}\setbox1=\hbox{#2}\ifdim
\wd1>\wd0\kern.5\wd1\kern-.5\wd0\fi
\copy0\kern-.5\wd0\kern-.5\wd1\copy1\ifdim\wd0>\wd1
\kern.5\wd0\kern-.5\wd1\fi}}
\def\ltap{\;\centeron{\raise.35ex\hbox{$<$}}{\lower.65ex\hbox{$\sim$}}\;}
\def\gtap{\;\centeron{\raise.35ex\hbox{$>$}}{\lower.65ex\hbox{$\sim$}}\;}

\newcommand{\newc}{\newcommand}
\newc{\qbar}{{\overline q}}
\newc{\Kahler}{K\"ahler }
\newc{\deltaGS}{\delta_{\rm GS}}

\begin{document}
\setlength{\unitlength}{1mm}
\preprint{
\vbox{\vspace*{2cm}
      \hbox{UCI-TR-2010-23}
      \hbox{October, 2010}
}}
\vspace*{3cm}

\title{Searches for Fourth Generation Charged Leptons}
\author{Linda M. Carpenter, Arvind Rajaraman and Daniel Whiteson}

\affiliation{Department of Physics and Astronomy  \\
   University of California, Irvine, CA 92697 \\
\vspace{1cm}}

\begin{abstract}

We study the production and decay of fourth generation leptons at the Large Hadron Collider (LHC).
We find that for charged leptons with masses under a few hundred GeV,
the dominant collider signal comes from the production through a W-boson of a
charged and neutral fourth generation lepton.  We present a sensitivity study
for this process in events
with two like-sign charged leptons and at least two associated jets.
We show that with $\sqrt{s}= 7$ TeV and 1 fb$^{-1}$ of data, the LHC can exclude fourth
generation charged leptons with masses up to 250 GeV.

\end{abstract}
\pacs{}

\maketitle

\section{Introduction}

One of the simplest 
possibilities for physics beyond the Standard Model
 is a fourth generation of fermions. Since these particles get masses from
  electroweak symmetry breaking, their masses must lie at the electroweak scale, and thus these
  particles are expected to be accessible to present and upcoming collider searches.
 Precision
  measurements constrain but do not exclude
  the parameter space for the fourth generation;
  with appropriate mass differences in the quark sector, a fourth generation is
   consistent with current electroweak precision data \cite{He:2001tp,Kribs:2007nz,Erler:2010sk}.

Direct searches at the Tevatron have constrained the fourth
generation $t'$ mass to be $\geq 335$ GeV~\cite{cdftp}, and the $b'$ mass is
constrained to be $\geq 385$ GeV~\cite{cdfbp}.  Many LHC search analyses for fourth generation quarks have also been performed (for example \cite{Ozcan:2008zz,Cakir:2008su,Holdom:2007ap}). However, experimental
limits on the corresponding fourth generation lepton sector from hadron colliders have not been widely explored
 until recently~\cite{ozcan:2009}.
 
 The most important  constraints on the fourth generation lepton sector currently come from LEP
 ~\cite{Achard:2001qw,Abulencia:2007ut}.
If the  fourth generation neutrinos can
decay to SM particles through charged current interactions,
they may be as light as 60 GeV~\cite{Carpenter:2010dt}.
 Similarly, bounds on unstable fourth generation charged
 leptons are $\sim 100$ GeV~\cite{Achard:2001qw}.
As we shall see, these bounds
 can be considerably improved at the LHC, even with just 1 fb$^{-1}$ of data.

 The
 masses in the lepton sector of the fourth generation
 are parametrized by independent Dirac
 masses for the fourth generation 
 charged leptons and neutrinos, as well
 as a Majorana mass for the right handed neutrino.  The presence of both Majorana
 and Dirac masses for neutrinos means that in general there are two independent neutrino mass
 eigenstates $N_1$ and $N_2$, with corresponding masses $M_1$ and $M_2$.  Furthermore, the charged lepton
 has an independent mass $M_L$. There are thus many possibilities
 for the mass spectrum in this sector.

 Here we consider the scenario where the lighter
neutrino $N_1$ is the lightest fourth generation lepton.  This neutrino
can then only decay through the process  $N_1 \rightarrow
W\ell$. Since the neutrino is a Majorana particle, it
can decay equally to  $W^-\ell^+$ and $W^+\ell^-$;
thus when a pair of fourth
generation leptons are produced, we expect the decay products to produce
like-sign dileptons -- a very distinctive signature -- half of the time.  This
can be used to significantly suppress backgrounds at hadron colliders.

In previous work~\cite{Rajaraman:2010ua,Rajaraman:2010wk}, 
the same-sign dilepton signature was  proposed and used to
study the sensitivity of fourth generation neutrino searches at hadron colliders.  However, these analyses assumed
that charged leptons were quite heavy, and only the fourth generation
neutrinos $N_1, N_2$ could be produced.
  In this work, we extend these analyses to include the charged lepton.

We will begin the next section by presenting the model that we consider, and
discussing the production
  and decay of
  fourth generation leptons
  at the LHC.  We find that the dominant production rate is that of
 charged lepton-neutrino through the process $qq'\rightarrow
  W\rightarrow LN_1$.
We also find that even for this very simple process the final state topology is quite
  complex, since  the leptons decay in a process $LN_1\rightarrow WWW\ell\ell$, where half the time the leptons are same sign.

 We then analyze the sensitivity of the LHC to fourth generation leptons in the
 channel with two like sign dileptons accompanied by two or more jets.
   Our sensitivity study shows that the 7 TeV LHC can
  exclude charged leptons with
  masses
  up to  250 GeV or better, even with just 1 fb$^{-1}$  of data.
  We conclude with a discussion of future directions.

\section{Fourth Generation Masses and Interactions}

 We will be following the notation of \cite{Katsuki:1994as}.

We are considering an extension to the standard model by a fourth generation
of fermions.   To
be completely general we include both Dirac
and Majorana masses for the left- and right-handed neutrinos.
The neutrino mass matrix may then be written as
\begin{eqnarray}
{\cal L}_m=-{1\over 2}\overline{(Q_R^c
N_R^c)}\left(\begin{array}{cc}0&m_D\\m_d&M\end{array}\right)
\left(\begin{array}{c}Q_R\\ N_R\end{array}\right)+h.c.
\end{eqnarray}
where $\psi^c=-i\gamma^2\psi^*$.
This theory contains two Majorana neutrinos $N_1, N_2$ with mass eigenvalues
\bea
 \nonumber M_1=-(M/2)+ \sqrt{m_D^2+{M^2/4}} \\ M_2=(M/2)+ \sqrt{m_D^2+{M^2/4}}
\eea
\noindent
In addition, there is a Dirac mass term for the fourth
generation lepton, $M_L\overline{L} E_R$, where
$L$ is the left handed doublet, and $E_R$ is the right handed singlet.

The  leptons couple to the gauge bosons through the interaction term
\bea
{\cal L}=gZ_\mu J^{\mu}+(gW_\mu^+ J^{\mu +} + c.c)
\eea
  where
\bea
 J^{\mu }
= {1\over 2\cos\theta_W}(-c^2_\theta\bar N_1\gamma^\mu \gamma^5N_1
-2is_\theta c_\theta\bar N_1\gamma^\mu N_2-s^2_\theta\bar N_2 \gamma^\mu\gamma^5N_2))
\\
J^{\mu +}=
c_i\overline{(c_\theta N_1-i s_\theta N_2)}\gamma^\mu l^i_L~ +{1\over\sqrt{2}} \overline{(c_\theta N_1-i s_\theta N_2)}\gamma^\mu L~~~~~~~~~~~~~~~~~~~
\eea
where $c_i$ are analogous to the CKM matrix elements.
 Here we have defined the mixing angle
\bea
\nonumber \tan\theta=M_1/m_D\label{mixingangle}
\eea

\begin{figure}
\begin{fmffile}{diag_0}
\begin{fmfchar*}(100,50)\fmfpen{thick}
 \fmflabel{$\ell$}{\ell1}
\fmf{plain}{v__N_11,\ell1}
 \fmflabel{$j$}{j1}
\fmf{plain}{v_W2,j1}
 \fmflabel{$j$}{j2}
\fmf{plain}{v_W2,j2}
\fmf{photon,label=$W $}{v__N_11,v_W2}
\fmf{plain,label=$N_1 $}{v_W1,v__N_11}
 \fmflabel{$\ell$}{\ell2}
\fmf{plain}{v__N_12,\ell2}
 \fmflabel{$j$}{j3}
\fmf{plain}{v__W1,j3}
 \fmflabel{$j$}{j4}
\fmf{plain}{v__W1,j4}
\fmf{photon,label=$W $}{v__N_12,v__W1}
\fmf{plain,label=$N_1 $}{v__L1,v__N_12}
 \fmflabel{$j$}{j5}
\fmf{plain}{v__Z1,j5}
 \fmflabel{$j$}{j6}
\fmf{plain}{v__Z1,j6}
\fmf{photon,label=$Z $}{v__L1,v__Z1}
\fmf{plain,label=$L $}{v_W1,v__L1}
\fmf{photon,label=$W $}{v_p_p1,v_W1}
\fmfright{\ell1,j1,j2,\ell2,j3,j4,j5,j6}
\fmf{plain}{p1,v_p_p1}
\fmflabel{$p$}{p1}
\fmf{plain}{p2,v_p_p1}
\fmflabel{$p$}{p2}
\fmfleft{p1,p2}
\end{fmfchar*}
\end{fmffile}
\caption{Feynman diagram of the production and decay of $L N_1$ pair.}
\label{fig:2pgm}
\end{figure}
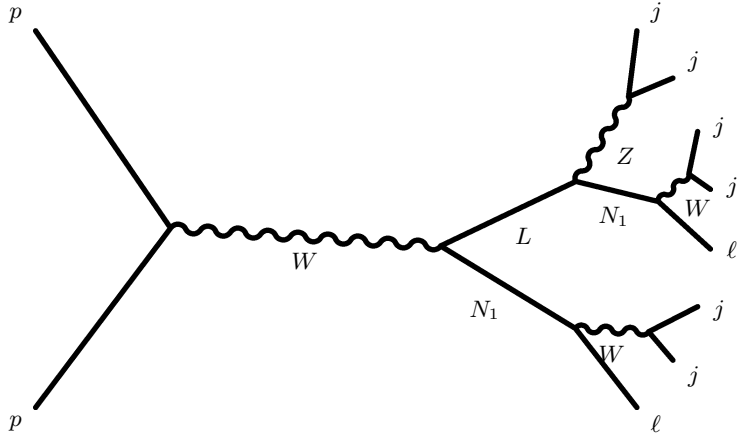

Since the parameters are not predicted theoretically, in principle either $N_1$ or $L$ could
be the lightest state (by construction $N_1$ is lighter than $N_2$).
In the following analysis we will consider the scenario where the
lightest fourth generation particle is $N_1$. It would be interesting to consider the
case where the charged lepton is the lightest new state; we leave this for future work.
 For the
moment we also assume that the mass of the heavier neutrino $N_2$ is greater than
that of the charged lepton;
later we will discuss the other situation when
$M_2<M_L$.

At hadron colliders, we can either pair produce
leptons through the processes $q\bar{q}\rightarrow Z\rightarrow N_iN_j$
and $q\bar{q}\rightarrow Z\rightarrow L^+L^-$, or we may produce a neutrino
and charged lepton through the process $q_i\bar{q}'_j\rightarrow W\rightarrow N_iL$.  At the LHC, we
have many more $W$'s than $Z$'s, due to the preponderance of u-quarks and d-antiquarks in a proton-proton
collision.  This implies that
if the charged lepton mass is
comparable to the neutrino mass,
the production rate of a charged lepton and
 neutrino through the $W$-boson
 will be much larger than the pair production rate
 through a $Z$.
 In fact, the typical production cross
sections for charged lepton-neutrino production
are found to be $10^{-1}-10^{-2}$ pb, which are greater than the pair
production rates
by almost
two orders of magnitude. We shall therefore ignore the pair production processes in the rest
of our analysis.

We begin by considering the case where the charged lepton
is produced along with the lighter neutrino
through the process  $pp\rightarrow W\rightarrow LN_1$.
Since the 
mixing between the fourth generation
and the first three generations is small (as required by precision experiments \cite{Chanowitz:2009mz}),
$L$ will decay dominantly through the
process
$L\rightarrow W N_1$.
$N_1$, on the other hand, can only decay
through the process 
$N_1\rightarrow \ell W$ where $\ell$ is one of the three leptons of the Standard Model.

The precise decay mode is controlled by the magnitude of the mixing angles.
For simplicity, we assume that the mixing with one of the three SM leptons  dominates,
and therefore that all $N_1 \rightarrow \ell W$ decays will proceed to the same flavor of SM
lepton.

The complete process that we consider is thus $pp\rightarrow LN_1
\rightarrow WN_1 N_1 \rightarrow WWW\ell\ell$, where the two charged
leptons are the same flavor, and in half the cases are of the same sign.
The $W$'s tend to decay hadronically; we therefore obtain two leptons
with multiple jets.
Note that though the underlying physics is quite simple, our process can yield an
eight body final state.

In general, we should also include production of $LN_2$. The production rate
for this process will be somewhat lower because of the higher $N_2$ mass. The
$N_2$ decays either as $N_2\rightarrow LW$ or $N_2\rightarrow N_1Z$; in either case, we
again get a signature of two
leptons with additional jets, which increases the sensitivity of our search. We
shall not include this production process in our analysis; our
bounds will therefore be conservative.

\begin{figure}[t]
\centerline{\includegraphics[width=7 cm]{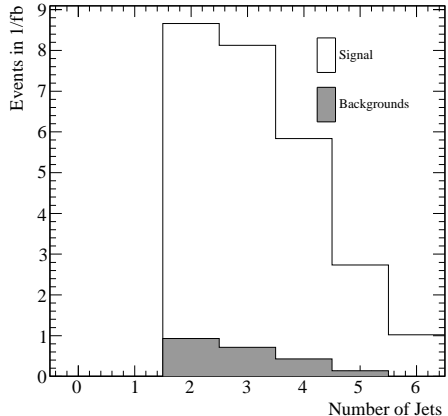}}
\caption{Distribution of expected jet multiplicity in 1 fb$^{-1}$ of
  data at the point $M_1=100, M_L=200$ after selection of like-sign
  dileptons and at least two jets.}
\label{fig:jetmult}
\end{figure}

\section{Sensitivity Analysis}

 We will focus on the $\ell^{\pm}\ell^{\pm}$ + multijets signature.
A histogram of jet multiplicities for a benchmark point is shown in Figure 2.
For example, for our
 benchmark point we expect 16 events per fb$^{-1}$ which contain 2 or more jets.

At the LHC, the largest backgrounds to the  $\ell^{\pm}\ell^{\pm}$ + multijets signature come
from $W\gamma$ or $WZ$ production or misidentified leptons either from
semi-leptonic $t\bar{t}$ decays or direct $W+$ jets production.
The LHC contains an additional process which was negligible at the
Tevatron,  $qq\rightarrow W^{\pm}W^{\pm}q'q'$, which directly produces
the $\ell^{\pm}\ell^{\pm}jj$ signature.  We calculate the size and kinematics of each contribution
using {\sc madgraph}~\cite{Alwall:2007st} and {\sc bridge}~\cite{Meade:2007js}, and
use {\sc pythia}~\cite{pythia} for
showering and a version of {\sc
  pgs}~\cite{pgs} tuned to describe the expected performance of the ATLAS detector.
   Figure 2  shows the expected background as a function of jet multiplicity.

Following~\cite{Rajaraman:2010ua}, we look at events with $N_{jet} > 2$ with like-sign
dileptons.  We impose the following
 cuts:

\begin{itemize}
\item{two isolated like-sign leptons of the same flavor }
\item{both lepton $p_T$ $>$ 25 GeV$/c$.}
\item{at least two jets with $E_T$ $>$ 20 GeV.}
\end{itemize}

\begin{figure}[t]
\begin{center}
\includegraphics[width=0.4\linewidth]{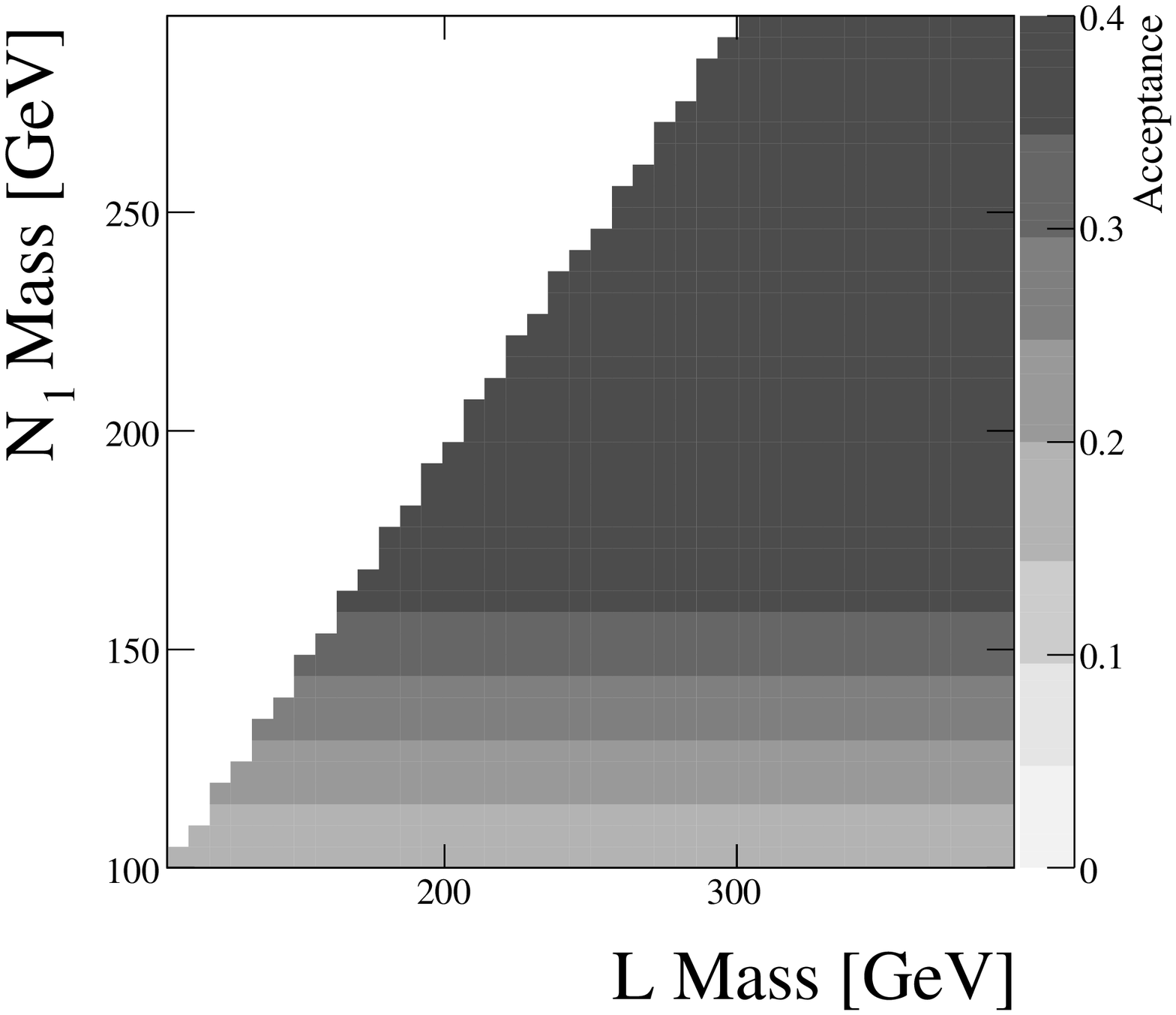}
\includegraphics[width=0.4\linewidth]{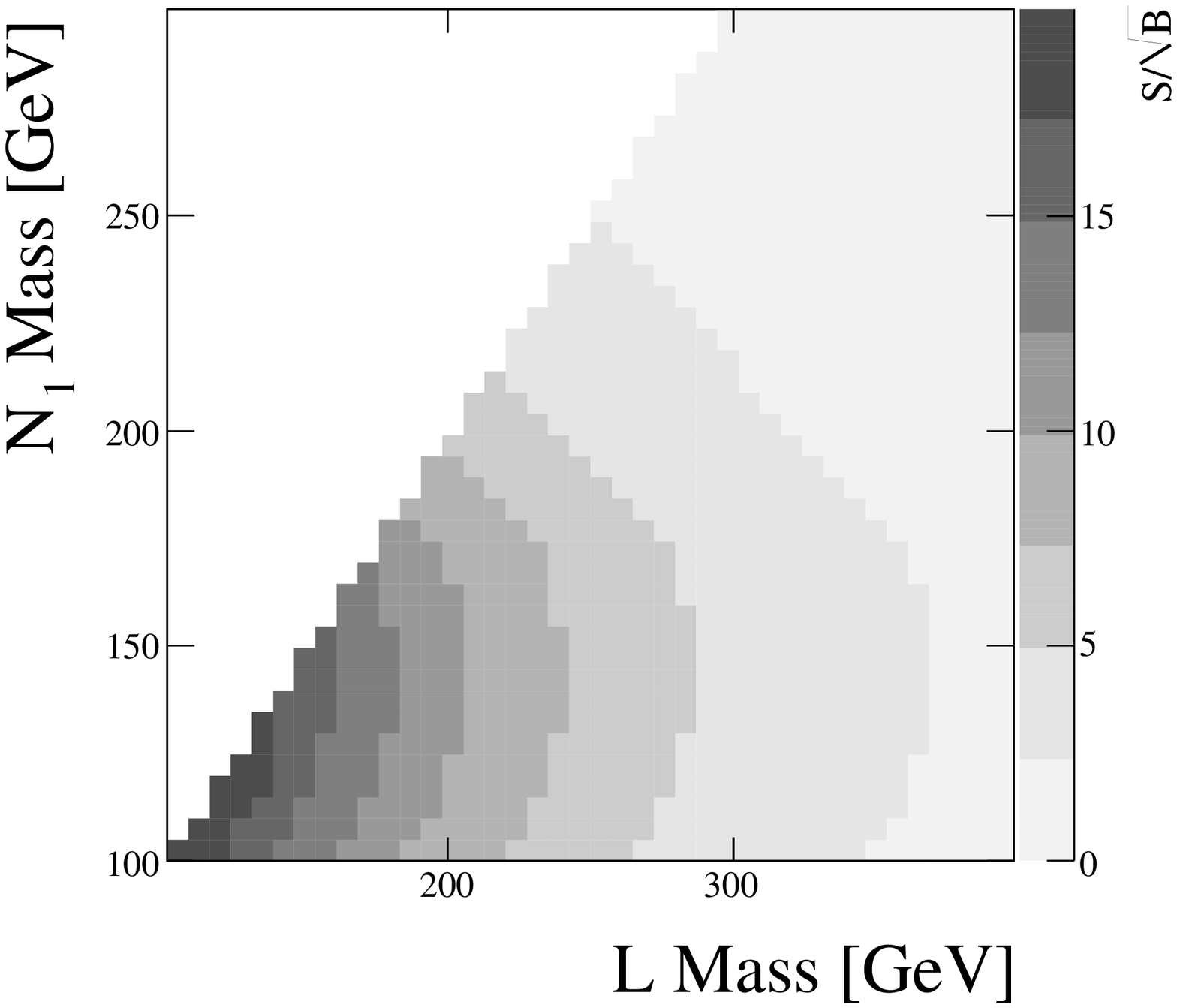}
\caption{  Left, search acceptance over the $M_{1}-M_{L}$
  plane. Right, a simple significance metric, $N_{signal} /
  \sqrt{N_{background}}$ in 1 fb$^{-1}$ of data.
}
\label{fig:reco}
\end{center}
\end{figure}

Figure 3 shows the calculated acceptances for these cuts in the $M_1-M_L$ mass plane.
The acceptances range from $0.15$ to $0.4$.
The efficiencies are relatively independent of the charged lepton mass;
they however drop as $N_1$ becomes less massive, because all the decay
products become soft.
Figure 3b gives an approximate figure of merit
$S/\sqrt{B}$ per fb$^{-1}$.  Note that the range of masses $M_1 > M_L$ is
excluded by assumption.

To extract more sensitivity and identify the mass scale,  we reconstruct the observed $N$ mass from the
$\ell jj$ objects, following the procedure in~\cite{Rajaraman:2010ua}.
We perform a binned likelihood fit in the reconstructed $N$
mass, and use the unified ordering scheme~\cite{feldcous} to construct
frequentist intervals.    We include an overall 100\% systematic
uncertainty on the background rate as well as an uncertainty that
describes our lack of understanding of the rate of radiation.

\begin{figure}[t]
\begin{center}

\includegraphics[width=0.4\linewidth]{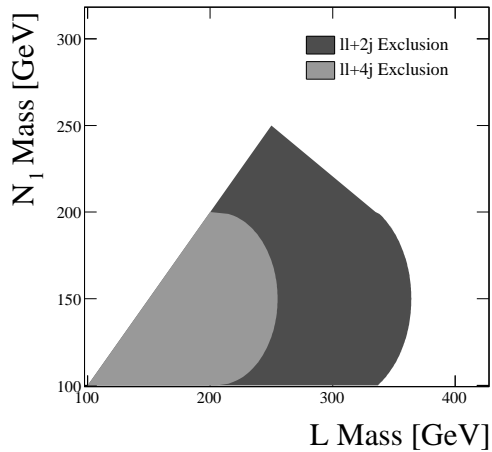}
\caption{ Expected exclusion in  $M_1-M_L$ mass plan for 1 fb$^{-1}$ for
  the dilepton+ 2j and dilepton + 4j selections.}
\label{fig:excl}
\end{center}
\end{figure}

The final expected exclusion ranges in the
$M_{1}-M_{L}$ plane are shown in Fig.~\ref{fig:excl}. We see that for 1 fb$^{-1}$ we may
discover or exclude
charged leptons of masses  up to 250 GeV.  For some values of $M_1$, charged leptons
  may be excluded up to masses of 320 GeV.

\section{Conclusions}

We have studied the fourth generation
leptonic sector, considering the case where both the
charged lepton $L$ and the fourth generation neutrinos $N_1,N_2$ are accessible
to colliders.
In this situation, the largest production cross section
of fourth generation particles is through $LN_1$ production.  The decay of this pair leads to a distinctive
final state topology 
of same-sign dilepton production in
association with multiple jets.  
We have shown that in this channel,  an LHC search at 7 TeV with 1 fb$^{-1}$ of
data
can exclude fourth generation charged leptons with masses up to 250 GeV. This is a significant
improvement on present constraints.

In the analysis above, we have assumed 
that the mass of $N_2$ is larger than that of the
charged lepton.  
However, the case that $N_2$ is lighter than $L$ is expected to be very similar. 
In that case, there may be sometimes be an $N_2$ in the $L$ decay chain,
$LN_1\rightarrow WN_2 N_1\rightarrow WZN_1N_1$. We still get
like sign dileptons from the $N_1$ decays, plus additional jets from the
extra $Z$.  We expect the efficiencies for this search to be the same or slightly better
than the one we have considered,
 as there will be more jets.  Our
bounds are only expected to be improved in this scenario.

It is also interesting to consider ways of distinguishing
this signal from other models of new physics with same-sign
dilepton signals, as for example pair production of $N_1$ in the situation
when $L$ is heavy.
One possibility is to
use the 
feature of the $LN$ final states 
that the
jet multiplicity falls off slowly.  For example,  Figure 2
shows that there are 16 event per fb$^{-1}$ with 2 jets or more, while there
are still 10 events per fb$^{-1}$ with 4 jets or more.   
Indeed, as shown in Figure 4, if
one requires a four jets plus like-sign dilepton signal instead of a
two jet plus dilepton signal, 
one can still exclude charged leptons
with masses
up to 200 GeV.  
 The presence
of many jets may therefore be a useful feature in distinguishing  the production of
$LN$ from other similar processes. It would be interesting to see how strongly the jet 
multiplicity constrains
the underlying process.

There are
several
further directions for study.
In particular, it would
be interesting to
consider 
the case where the charged lepton is the lightest of the fourth generation
leptons.
Another possibility is for the lightest neutrino $N_1$ to be stable.
In this scenario,
instead of dileptons, we
obtain signatures with a large amount of missing energy, which  may be challenging
to observe. We hope to return to these analyses in future work.

\section{ Acknowledgments}

We thank  Tim Tait and Gokhan Unel for helpful discussions.
This work was supported in part by the DOE under grant number
DE-FG03-92ER40689 and in part by the NSF under grant number PHY-0970173.

\end{document}